\documentclass[twocolumn,aps,showpacs,amssymb,floatfix]{revtex4}
\usepackage{graphicx}

\usepackage{epsfig}

\newcommand{\be}{\begin{equation}}
\newcommand{\ee}{\end{equation}}
\newcommand{\bea}{\begin{eqnarray}}
\newcommand{\eea}{\end{eqnarray}}
\newcommand{\beq}{\begin{eqnarray}}
\newcommand{\eeq}{\end{eqnarray}}

\begin{document}

\title{Baryon structure from Lattice QCD}

\author{%
      C. Alexandrou$^{1,2}$\email{alexand@ucy.ac.cy}%
}

\address{%
$^1$~Department of Physics, University of Cyprus, P.O. Box 20357, Nicosia, CY-1678 Cyprus\\
$^2$~Computation-based Science and Technology Research Center, Cyprus Institute, 20 Kavafi Street
Nicosia 2121, Cyprus\\
}

\begin{abstract}
We present recent lattice results on the baryon spectrum,  nucleon electromagnetic and axial form factors, nucleon to $\Delta$ transition
form factors as well as the $\Delta$ electromagnetic form factors.
The masses of the low lying baryons and the  nucleon form factors are calculated using two degenerate flavors of twisted
mass fermions down to pion mass of about 270 MeV. We compare 
to the results of other collaborations. 
The nucleon to $\Delta$ transition
and  $\Delta$ form factors are calculated in a hybrid scheme, which
uses  staggered sea quarks and domain wall valence quarks. The dominant dipole nucleon to $\Delta$ transition
form factor is also evaluated using dynamical domain wall fermions.
The transverse density distributions of the $\Delta$ in the infinite momentum frame
are extracted using the form factors determined from lattice QCD.  
\end{abstract}


\maketitle

\section{Introduction}
\vspace*{-0.5cm}
During the last five years we have seen tremendous progress in dynamical lattice simulations using
a number of different fermion discretization schemes with quark masses  reaching closer to the physical pion mass.
Many collaborations are contributing to this progress. The European Twisted Mass Collaboration (ETMC) 
is using twisted mass fermions (TMF), which provide an attractive  formulation of lattice QCD that
allows for automatic ${\cal O}(a)$ improvement, infrared regularization 
of small
eigenvalues and fast dynamical 
simulations~\cite{Frezzotti:2000nk,Frezzotti:2003ni}.
Automatic  
 ${\cal O}(a)$ improvement is obtained by tuning only one parameter
requiring no further improvements on the operator level. A
 drawback of twisted mass fermions is the ${\cal O}(a^2)$ breaking of isospin
symmetry, which is only restored in the continuum limit.  
In the baryon sector it has been shown that this isospin breaking is consistent with zero
within our
statistical accuracy  by evaluating the mass difference
between $\Delta^{++}(\Delta^-)$ and $\Delta^{+}(\Delta^0)$~\cite{Alexandrou:2008tn,Alexandrou:2007qq}.
This is in agreement with a theoretical 
analysis~\cite{Frezzotti:2007qv,Shindler:2007vp} that shows
potentially large ${\cal O}(a^2)$ flavor breaking effects to
 appear in the $\pi^0$-mass
but to be suppressed in  other quantities.
A number of collaborations, as for example
 QCDSF~\cite{Gockeler:2007rm}, PACS-CS~\cite{Aoki:2008sm}, 
BMW~\cite{Durr:2008zz} and CERN~\cite{DelDebbio:2006cn} are using improved 
Clover fermions for their simulations. 
It is worth mentioning that PACS-CS has simulations very close to the physical pion mass albeit in a small
volume,  whereas the Wuppertal group recently calculated  meson masses and
the decay constants using $N_F=2+1$ configurations simulated at the physical pion mass ~\cite{Aoki:2009sc}.
 A number of groups adopted a hybrid approach to compute hadronic matrix elements
 taking advantage of the efficient simulation and availability of staggered sea fermions produced
by the MILC collaboration~\cite{Aubin:2004wf}  and the chiral symmetry
of domain wall fermions. The Lattice Hadron Physics
Collaboration (LHPC) has been particularly active in producing  results on a number of key 
observables~\cite{Hagler:2007xi,Alexandrou:2009hs}, some of which will be discussed in Sections IV and V.
A very promising recent development is the simulation of dynamical chiral fermions 
using large volumes and at small enough pions masses. The RBC-UKQCD collaboration
is generating gauge configurations using $N_F=2+1$ domain wall fermions (DWF)~\cite{Boyle:2007fn},
whereas the JLQCD Collaboration
is producing dynamical configurations with two flavors of overlap fermions~\cite{Aoki:2008tq}.
Most of the current simulations  are done using volumes of spatial length $L$ such that  $ m_\pi L>3.5$ to
keep finite volume effects small. The fact that simulations in the chiral regime are possible is to a large extend 
due to algorithmic improvements that  
yield better scaling behavior as the physical pion mass is approached. For a discussion on the
scaling and a comparison among  the different fermion discretization schemes see Ref.\cite{Jansen:2008vs}.

\section{Hadron spectrum}
\begin{center}
\begin{figure}
\includegraphics[width=7.5cm,height=12cm]{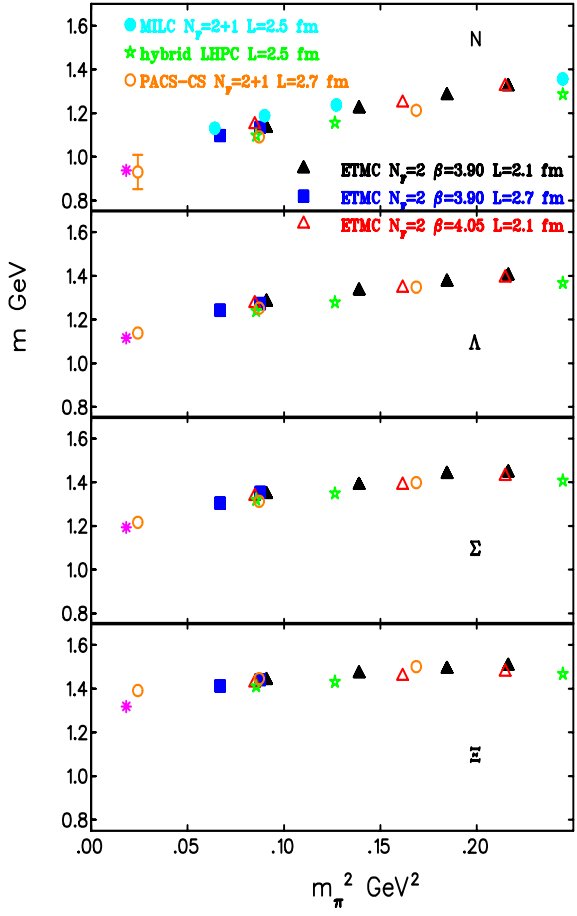}
\caption{\label{fig:octet} Comparison of masses for the low lying octet 
baryons. Results using $N_F=2$ TMF are shown by the 
filled (black) triangles for $L=2.1$~fm and (blue) squares for $L=2.7$~fm with $a=0.089$~fm and with the open (red) triangles  for $L=2.1$~fm and $a=0.070$~fm. Results with the hybrid action are shown with the (green) asterisks for $a=0.124$~fm and results using $N_F=2+1$ 
Clover fermions with
the open (orange) circles and $a=0.0907$~fm. 
For the nucleon we also show results
using $N_F=2+1$ staggered fermions (filled (light blue) circles).
The physical masses are shown by the (purple) star. }
\end{figure}
\end{center}

\begin{center}
\begin{figure}
\includegraphics[width=7.5cm,height=12cm]{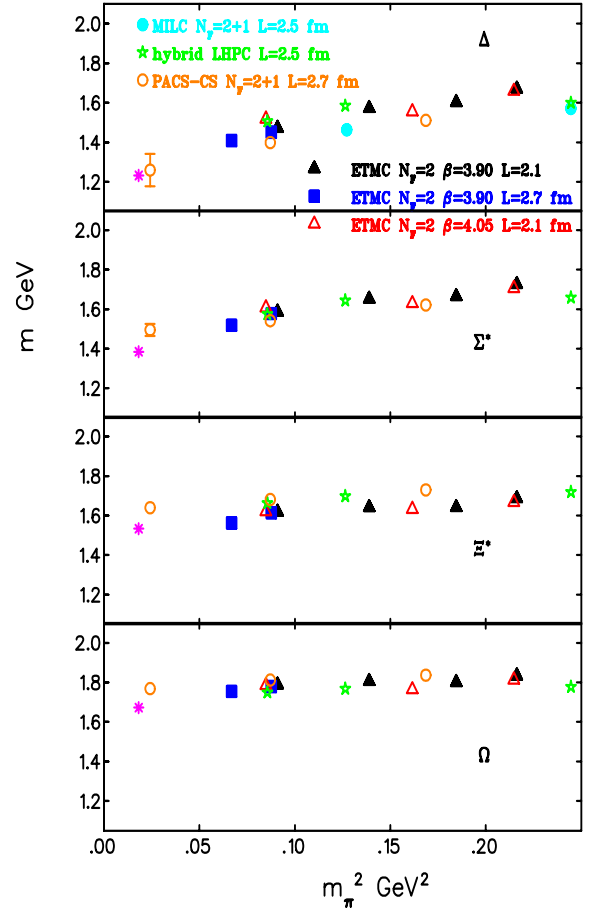}
\caption{\label{fig:decuplet} Comparison of masses for the low lying decuplet baryons.The notation is the same as that of Fig.~\ref{fig:octet}. }
\end{figure}
\end{center}

The masses of the lowest lying hadrons of a given set of quantum numbers are
readily calculated by computing the two-point function at zero momentum:
$C_h(t)=\sum_{\bf x}\langle 0 |J_h({\bf x},t)J_h^\dagger(0)|0 \rangle $. 
Choosing good interpolating fields and 
applying smearing techniques ensure ground state dominance at short time
 separation $t$ so that  gauge noise is kept small~\cite{Alexandrou:2006ru}.
In Figs.~\ref{fig:octet} and \ref{fig:decuplet} we compare recent
results on the low lying baryon spectrum using dynamical twisted mass~\cite{Alexandrou:2008tn,Drach:2009dh}
and clover fermions~\cite{Aoki:2008sm} and within the hybrid approach~\cite{WalkerLoud:2008bp}
(staggered sea and domain wall valence quarks).
The level of agreement  of lattice QCD 
results using a variety of fermion discretization schemes 
  seen in Figs.~\ref{fig:octet} and \ref{fig:decuplet}  
before taking the continuum limit or other lattice artifacts into account 
is quite impressive. Small discrepancies seen mainly in
the decuplet masses can be attributed to lattice artifacts and a systematic analysis of these effects 
is performed by each collaboration before extracting the final continuum values. In
particular results using  staggered fermions may suffer the most from cut-offs effects
since the lattice used is  rather coarse  as compared to those using twisted mass and Clover fermions which have lattice spacings smaller than 0.1~fm.

Having reliable methods to extract the masses of the low lying hadrons one
can  investigate techniques for the extraction of the masses of excited states.
A number of approaches exist. A commonly used method is based on the variational
approach~\cite{var}: 
For a given $N\times N$ correlator matrix  
$C_{kn}(t)=\langle 0 |J_k(t)J_n^\dagger(0)|0 \rangle $
 one defines the $N$ principal correlators with $\lambda_k(t,t_0)$
 as the eigenvalues of $C(t_0)^{-1/2} C(t) C(t_0)^{-1/2}$, 
 where $t_0$ is small. Since
\be \lim_{t\rightarrow \infty} \lambda_k(t,t_0)=e^{-(t-t_0)E_k}
\left(1+e^{-t\Delta E_k}\right), \quad k=1,\dots, N
\ee
 the $N$ principal effective masses tend (plateau) to the N 
lowest-lying stationary-state energies of the hadrons with the same quantum numbers.
 It is crucial to use very good operators so noise does not swamp signal
and  construct 
 spatially extended operators using smearing of the quark fields as well 
as applying link variable smearing. The use of a 
 large set of appropriately constructed operators is also very important.
Despite recent calculations using this method~\cite{var_results} 
the issue of the ordering of the Roper resonance as compared to the negative parity
partner of the nucleon still remains unresolved.
Maximum entropy methods have also be developed for the analysis of hadron 
two-point correlators and recent results can be found in Ref.~\cite{MEM}.
A new method that relies solely on 
$\chi^2$-minimization with an unbiased evaluation of errors
can be applied to extract the masses of the states on which the 
two-point correlator is sensitive on~\cite{AMIAS}. 
This method was applied to extract  the excited states of the nucleon using
local correlators that are easily produced  in lattice simulations.
For this study two interpolating fields are considered:
\beq
 J_N(x)&=&\epsilon^{abc} (u_a C\gamma_5 d_b^T)u_c\nonumber \\
J^\prime_N(x)&=&\epsilon^{abc} (u_a^T C d_b)\gamma_5 u_c. 
\label{interpolating fields}
\eeq
 \begin{center}
\begin{figure}
\includegraphics[width=7.5cm,height=7.5cm]{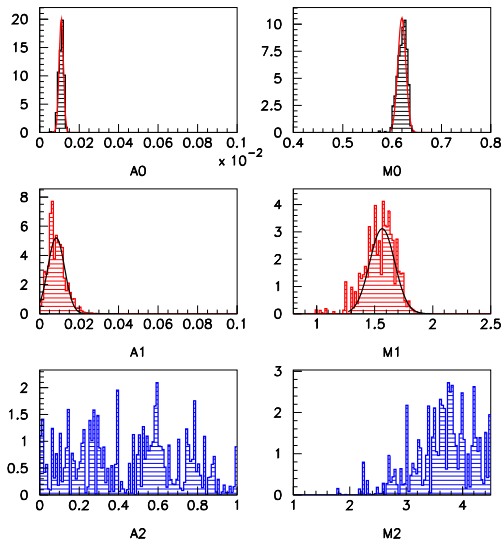}
\caption{\label{fig:N pos}Probability distributions for the amplitudes and masses in lattice
units extracted from
 local correlators using $N_F=2$ Wilson fermions  at pion mass
500~MeV  on a lattice of spatial length 1.8~fm at $\beta=6.0$ using $J_N$.}
\end{figure}
\end{center}
 \begin{center}
\begin{figure}
\includegraphics[width=7.5cm,height=7.5cm]{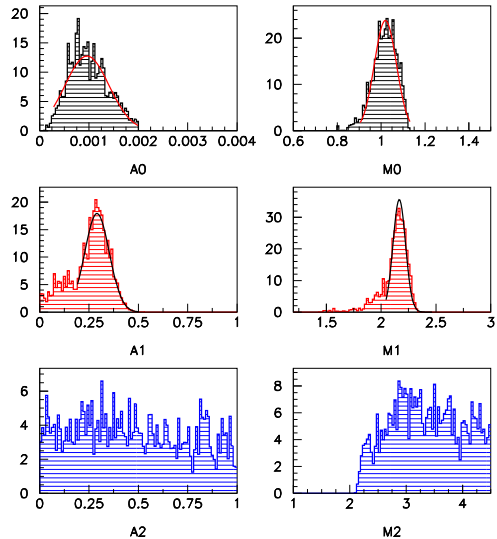}
\caption{\label{fig:N pos prime} As in Fig.~\ref{fig:N pos} but 
using the interpolating field $J_N^\prime$.}
\end{figure}
\end{center}
As can be seen from the histograms shown in Figs.~\ref{fig:N pos} and
\ref{fig:N pos prime}, one clearly identifies the first excited state in the
positive parity channel of the nucleon using rather
low quality data.
In addition, 
we observe that  the state of lowest mass that is present in the mass spectrum of
the correlator computed with $J_N$ is absent when using $J^\prime_N$.
Instead the correlator with $J_N^\prime$ has a lowest state 
that does not show up
when using $J_N$.  The conjecture is that this state is the Roper.

\section{Form factors}
To extract information on hadron structure one needs to calculate coupling 
constants, such as the nucleon axial charge $g_A$, the $\pi N$ and $\pi N\Delta$ coupling
constants, form factors, 
moments of parton distribution functions
and  generalized form functions.
In order to compute these quantities we need to calculate the relevant three-point functions, which,
in addition to the forward propagator needed for the calculation of the masses, require the evaluation of the
sequential propagator. The three-point function, related to
the matrix element of the operator ${\cal O}$ between hadron states $|h^\prime>$ and $|h>$, is given by

\beq
 && \hspace*{-1cm}\langle G^{h^\prime {\cal O} h}(t_2, t_1 ;
{\bf
p}^{\;\prime}, {\bf p}; \Gamma) \rangle =
  \sum_{{\bf x}_2, \;{\bf
x}_1} \exp(-i {\bf p}^{\;\prime} \cdot {\bf x}_2 ) \exp(+i {\bf q} \cdot {\bf x}_1 ) \; 
\nonumber \\ &&\langle \;\Omega \; | \;
\Gamma^{\beta \alpha}T\left[J^{\alpha}_{h^\prime} ({\bf x}_2,t_2) {\cal O}({\bf
x}_1,t_1) \bar{J}^{\beta}_h ({\bf 0},0) \right] \; | \;\Omega
\;\rangle   \; , 
\label{3pt}
\eeq 
where for ${\cal O}$ we consider the electromagnetic and axial currents.
We use sequential inversions through the sink, which allows us to obtain the
three-point function for any momentum transfer ${\bf q}$ and
operator insertion but fixes the quantum numbers of the initial and final baryons.

\subsection{Nucleon Electromagnetic form factors}
The elastic nucleon electromagnetic form factors  are fundamental 
quantities
characterizing important features of  neutron and  proton structure
that include their size, charge distribution and magnetization.
 An accurate  determination 
of these quantities in lattice QCD is timely and important
because of a new generation of precise experiments.
The matrix element of interest is 
\small
\beq 
\langle N(p',s')|A_\mu^3|N(p,s)\rangle=  \Bigg(\frac{
            m_N^2}{E_N({\bf p}')E_N({\bf p})}\Bigg)^{1/2} \nonumber \\
         \bar{u}(p',s') \Bigg[ \gamma_\mu F_1(q^2) 
+  \frac{i\sigma_{\mu\nu}q^\nu}{2m_N} F_2(q^2)\Bigg] \; u(p,s),
\label{NjN}
\eeq
\normalsize
where  $p(s)$ and $p'(s')$ denote initial and final momenta (spins) and 
$ m_N$ is the nucleon mass, 
$F_1(0)=1$ for the proton and
$F_2(0)$ measures the anomalous magnetic moment. These form factors are connected to the  
electric, $G_E(q^2)$, and magnetic, $G_M(q^2)$, Sachs form factors by the relations
\beq
G_E(q^2)&=& F_1(q^2) + \frac{q^2}{(2m_N)^2} F_2(q^2)\nonumber \\
G_M(q^2)&=& F_1(q^2) + F_2(q^2) \quad .
\label{Sachs ff}
\eeq
To extract the nucleon matrix element 
from lattice measurements, we calculate, besides the three 
point function 
$G^{Nj^\mu N} (t_2, t_1 ; {\bf p}^{\;\prime}, {\bf p};\Gamma )$,
the nucleon  two-point function, $ G^{NN}(t,{\bf p})$, 
and look for a plateau in  
the large Euclidean
time behavior of the ratio 
\small
\beq
R (t_2, t_1; {\bf p}^{\; \prime}, {\bf p}\; ; \Gamma ; \mu) &=&
\frac{\langle G^{Nj^\mu N}(t_2, t_1 ; {\bf p}^{\;\prime}, {\bf p};\Gamma ) \rangle \;}{\langle G^{N N}(t_2, {\bf p}^{\;\prime};\Gamma_4 ) \rangle \;} \> \nonumber \\
&\>& \hspace*{-3.8cm}\biggl [ \frac{ \langle G^{N N}(t_2-t_1, {\bf p};\Gamma_4 ) \rangle \;\langle 
G^{NN} (t_1, {\bf p}^{\;\prime};\Gamma_4 ) \rangle \;\langle 
G^{N N} (t_2, {\bf p}^{\;\prime};\Gamma_4 ) \rangle \;}
{\langle G^{N N} (t_2-t_1, {\bf p}^{\;\prime};\Gamma_4 ) \rangle \;\langle 
G^{N N} (t_1, {\bf p};\Gamma_4 ) \rangle \;\langle 
G^{N N} (t_2, {\bf p};\Gamma_4 ) \rangle \;} \biggr ]^{1/2} \nonumber \\
&\;&\hspace*{-1cm}\stackrel{t_2 -t_1 \gg 1, t_1 \gg 1}{\Rightarrow}
\Pi({\bf p}^{\; \prime}, {\bf p}\; ; \Gamma ; \mu) \; .
\label{R-ratio}
\eeq
\normalsize
where
\small
\be
\langle G^{NN} (t, {\bf p} ; \Gamma) \rangle =  \sum_{{\bf x}}
e^{-i {\bf p} \cdot {\bf x} } \; \Gamma^{\beta \alpha}\;\langle \Omega |\;T\;J^{\alpha}({\bf x},t) 
 \bar{J}^{\beta} ({\bf 0},0)  
\; |  \Omega\;\rangle .
\ee
\normalsize
We use the lattice conserved   electromagnetic current,   $j^\mu (x)$,
symmetrized on site $x$
and projection matrices for the Dirac indices
\be
\Gamma_i = \frac{1}{2}
\left(\begin{array}{cc} \sigma_i & 0 \\ 0 & 0 \end{array}
\right) \;\;, \;\;\;\;
\Gamma_4 = \frac{1}{2}
\left(\begin{array}{cc} I & 0 \\ 0 & 0 \end{array}\right) \quad.
\ee
Throughout this work we use kinematics where the final nucleon state
 is produced at rest
and therefore ${\bf q}={\bf p}^{\prime}-{\bf p}=-{\bf p}$.  
For the polarized matrix element one can construct
an optimal linear
combination for the nucleon sink, which in Euclidean time is given by 
\small
\beq  S_m({\bf q}; i)= \sum_{k=1}^3\Pi (-{\bf
q}\; ;
\Gamma_k ;\mu=i) =  \frac{C}{2m_N} \biggl\{ (p_2-p_3)\delta_{1,i} \nonumber \hspace{-0.5cm} \\
 + (p_3-p_1)\delta_{2,i} + (p_1-p_2)\delta_{3,i} \biggr\}
G_M(Q^2) 
\label{GM optimal}
\eeq 
\normalsize
with $Q^2=-q^2$. This construction
provides the maximal set of
lattice measurements from which  $G_M(Q^2)$
can be extracted requiring one sequential inversion.
No such
improvement is necessary for the unpolarized matrix elements
given by
\be \Pi ( {\bf 0}, -{\bf q}\; ; \Gamma_4 \; ; \mu=i )  = C
\frac{q_i}{2 m_N} \; G_E (Q^2) 
\label{GE123}
\ee
and
\be \Pi ( {\bf 0}, -{\bf q}\; ; \Gamma_4 \; ; \mu = 4)  = C
\frac{E_N +m_N}{2 m_N} \; G_E (Q^2)  \; ,
\label{GE4}
\ee
which yield $G_E(Q^2)$ with an additional sequential inversion.
$C=
\sqrt{\frac{2 m_N^2}{E_N(E_N + m_N)}}$ is a factor due to the
normalization of the lattice states. 
 \begin{center}
\begin{figure}
\includegraphics[width=7.5cm,height=5cm]{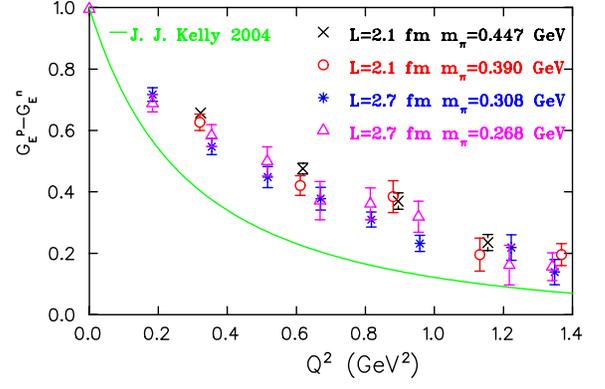}
\caption{\label{fig:GE} Nucleon isovector electric form factor using $N_F=2$ TMF.}
\end{figure}
\end{center}
\begin{center}
\begin{figure}
\includegraphics[width=7.5cm,height=5cm]{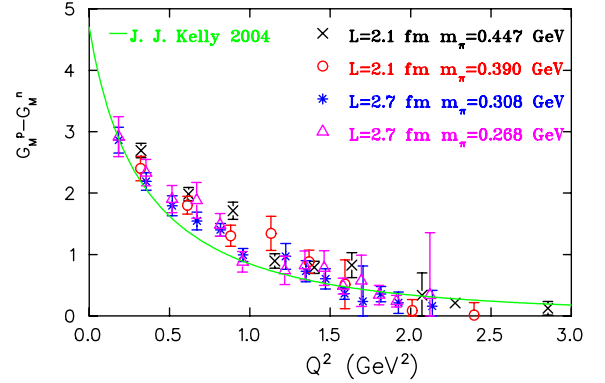}
\caption{\label{fig:GM} Nucleon isovector magnetic form factor using $N_F=2$ TMF.}
\end{figure}
\end{center}
Besides using an optimal nucleon source, the other
important ingredient in the extraction of the form factors
 is to take into
account simultaneously
in our analysis  all the lattice momentum vectors that contribute to a given 
$Q^2$. This is done by solving the overcomplete set of equations
\be
P({\bf q};\mu)= D({\bf q};\mu)\cdot F(Q^2) 
\label{overcomplete}
\ee
where $P({\bf q};\mu)$ are the lattice measurements of the ratio
given in Eq.~(\ref{R-ratio}) having statistical errors
$w_k$ and using the different sink types,
$F =  \left(\begin{array}{c}  G_{E} \\
                                    G_M \end{array}\right)$
and $D$ is an $M\times 2$ matrix which depends on 
kinematical factors with $M$ being the number of current 
directions and momentum vectors contributing to 
a given $Q^2$. We extract the form factors by 
minimizing 
\be
\chi^2=\sum_{k=1}^{N} \Biggl(\frac{\sum_{j=1}^2 D_{kj}F_j-P_k}{w_k}\Biggr)^2
\ee
using the singular value decomposition of $D$.
The analysis described in this Section to extract $G_{E}(Q^2)$ and $G_M(Q^@)$
 is also applied to the analysis of all form factors presented in this work. 

The $\gamma\,N \rightarrow N$  transition contains isoscalar photon
contributions. This means that disconnected loop diagrams also
contribute. These are generally difficult to
evaluate accurately since the all-to-all quark
propagator is required. In order to avoid
disconnected diagrams, we calculate the isovector form factors.
 Assuming $SU(2)$ isospin
symmetry, it follows that 
$
\langle \; p \,| (
\frac{2}{3}\bar{u} \gamma^{\mu}u - \frac{1}{3}\bar{d} \gamma^{\mu}
d ) | p \rangle  - \langle \; n | ( \frac{2}{3}\bar{u}
\gamma^{\mu}u - \frac{1}{3}\bar{d} \gamma^{\mu} d ) | n \rangle \;
\nonumber = \langle \; p \, | ( \bar{u}
\gamma^{\mu} u - \bar{d} \gamma^{\mu} d ) | p \rangle$ and 
 therefore by calculating  the proton three-point
function related to the matrix element of  the right hand side of the above relation we obtain
 the {\it isovector } nucleon form
factors $G^p_E (q^2)\, - G^n_E(q^2)$ and
 $ G^p_M(q^2)- G^n_M (q^2)$. 

\begin{center}
\begin{figure}
\includegraphics[width=7.5cm,height=5cm]{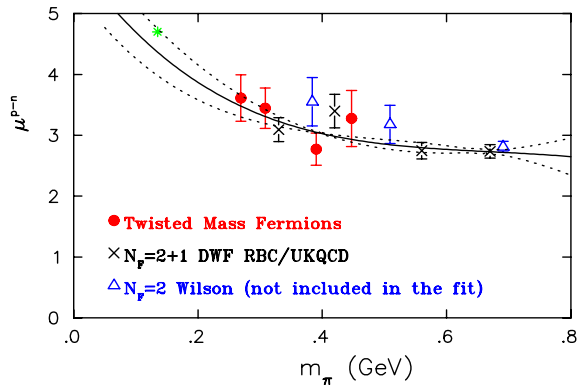}
\caption{\label{fig:moment} Nucleon magnetic moment using $N_F=2$ TMF (filled circles),
 $N_F=2+1$ DWF~[31]
(crosses) and $N_F=2$ Wilson fermions~[32]
 (open triangles). The physical value is shown by the star.}
\end{figure}
\end{center}

The results for the isovector electric and magnetic form factors using $N_F=2$ twisted mass
fermions are shown in Figs.~\ref{fig:GE} and \ref{fig:GM}~\cite{Alexandrou:2008rp}. The lattice
results on the electric form
factor fall off  slower as compared to 
a parametrization of the experimental data~\cite{Kelly:2004hm} shown by the solid line,
whereas  the magnetic form factor is closer to experiment. 
This is consistent with  recent high accuracy results obtained
 by the LHP Collaboration~\cite{Bratt:2008uf} and 
 the RBC-UKQCD Collaboration~\cite{Ohta:2008kd}.
Fitting the magnetic form factor to a dipole form we extract $G_{M}(0)$, which determines
the anomalous magnetic moment. We show its dependence on the pion mass in Fig.~\ref{fig:moment}.
Using
chiral effective theory with explicit nucleon and $\Delta$ degrees of 
freedom to one-loop order the isovector anomalous magnetic moment~\cite{chiral}, 
the  Dirac and Pauli radii can be extrapolated to the physical point~\cite{chiral,QCDSF,Alexandrou:2006ru}.
There are three fit parameters for the magnetic moment and the best fit to the twisted mass data with the associated error band is shown in Fig.~\ref{fig:moment}. 
 Multiplying the Pauli
radius squared with the magnetic moment yields an expression with only one-parameter like the Dirac 
radius that can shift the curves but does not affect their slopes.
 As can be seen the physical magnetic moment is within the error band whereas for the radii  
results closer to the physical point are needed to check
the predicted slope.

\begin{center}
\begin{figure}
\includegraphics[width=7.5cm,height=9cm]{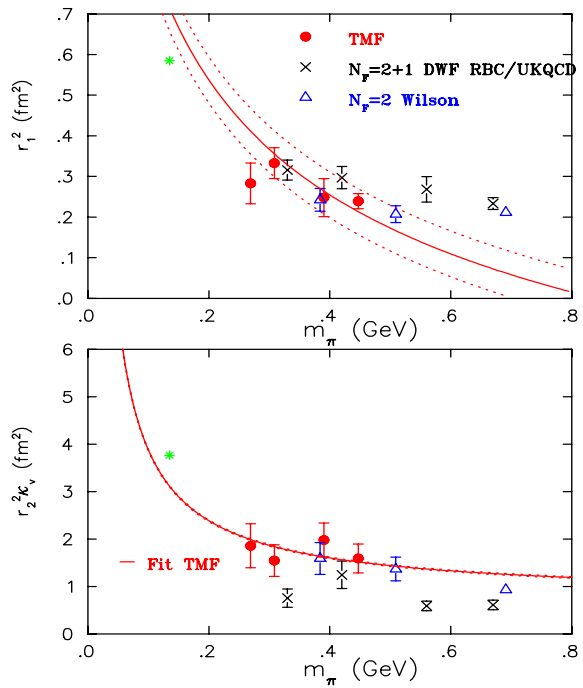}
\caption{\label{fig:radii} The Dirac radius squared (top) and Pauli radius squared
multiplied by the magnetic moment (bottom). The notation is the same
as that of Fig.~\ref{fig:moment}.}
\end{figure}
\end{center}

\subsection{Nucleon axial form factors}
The matrix element  of the weak axial vector current between nucleon states
can be written as
\vspace*{-0.5cm}

\beq 
\langle N(p',s')|A_\mu^3|N(p,s)\rangle= i \Bigg(\frac{
            m_N^2}{E_N({\bf p}')E_N({\bf p})}\Bigg)^{1/2} \nonumber \\
            \bar{u}(p',s') \Bigg[
            G_A(q^2)\gamma_\mu\gamma_5 
            +\frac{q_\mu \gamma_5}{2m_N}G_p(q^2) \Bigg]\frac{\tau^3}{2}u(p,s)
\label{NN axial}
\eeq 
where the axial isovector current 
 $A^3_\mu=\bar{\psi}(x)\gamma_\mu\gamma_5\frac{\tau^3}{2}\psi(x)$.
\begin{center}
\begin{figure}
\includegraphics[width=7.5cm,height=5cm]{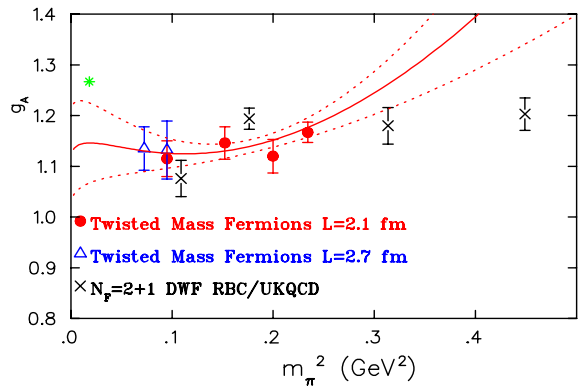}
\caption{\label{fig:gA} The nucleon axial charge using $N_F=2$ TMF and  $N_F=2+1$ DWF.}
\end{figure}
\end{center}

\begin{center}
\begin{figure}
\includegraphics[width=7.5cm,height=9cm]{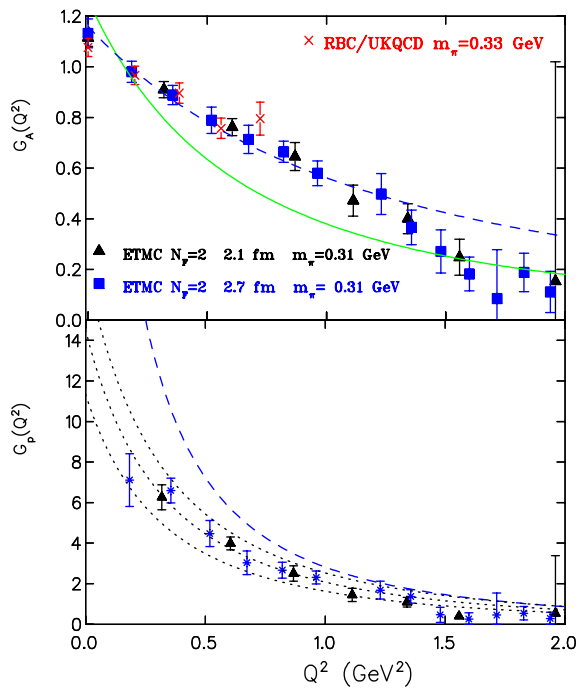}
\caption{\label{fig:axial} Top: The nucleon axial form factors 
$G_A(Q^2)$ for $N_F=2$ TMF and  $N_F=2+1$ DWF.
The dashed line is a dipole fit to the lattice data whereas the solid line
to experiment. Bottom:  $G_p(Q^2)$ for $N_F=2$ TMF. 
The dashed line is predicted from
$G_A(Q^2)$ and Eq.~(\ref{pole dominance}). The dotted lines is the
 best fit 
with the associated error band.}
\end{figure}
\end{center}
 Having computed the 
nucleon electromagnetic form factors we can obtain the axial 
ones with no additional 
inversions\cite{Alexandrou:2007zz, Alexandrou:2008rp}. 
The advantage here is that only the connected diagram contributes.
 In addition
at zero momentum transfer we obtain the nucleon axial charge $g_A$,
 a quantity that  is very accurately
measured experimentally. We show in Fig.~\ref{fig:gA} results obtained using $N_F=2$
twisted mass~\cite{Alexandrou:2008rp} and domain wall fermions~\cite{Yamazaki:2008py}.
The leading one-loop chiral perturbation theory result for $g_A$ in the small scale expansion~\cite{Hemmert:2003cb}
 can be used to
extrapolate lattice results to the physical point. Making a three-parameter fit to the twisted mass results we
obtain the solid curve shown in Fig.~\ref{fig:gA} together with the error band determined by allowing the
fit parameters to vary within a $\chi^2$ increase by one unit from the minimum. Note that this error band does not 
include uncertainties in the fixed parameters. We obtain, at
 the physical point, a value with a large error that is lower than
 the experimental value by an amount slightly larger than one standard deviation.
Results closer to the physical pion mass are 
needed to reduce the error due to the chiral extrapolation. 

 The $Q^2$-dependence of the  nucleon axial form factors 
$G_A(Q^2)$  and $G_p(Q^2)$  using $N_F=2$ twisted mass fermions is shown in Fig.~\ref{fig:axial}.
Our results for $G_A(Q^2)$ are in agreement with those obtained using  $N_F=2+1$ domain wall fermions 
at a comparable value of the pion mass. The $Q^2$-dependence of $G_A(Q^2)$  can be well described
by a dipole Ansatz 
$g_0/(Q^2/m_A^2+1)^2$ as shown by the dashed line.
This is what is usually 
used to describe experimental data for $G_A(Q^2)$
 where a value of  $m_A\sim 1.1$~GeV
is extracted for the axial mass. However the axial mass $m_A$ extracted from the lattice data
is larger resulting in a slower fall off as compared to experiment shown by the solid line.
 Assuming pion pole dominance $G_p(Q^2)$ can be obtained in terms of $G_A(Q^2)$ as
\be
G_p(Q^2)=\frac{4m_N^2/m_\pi^2}{1+Q^2/m_\pi^2}\> G_A(Q^2) .
\label{pole dominance}
\ee
In Fig.~\ref{fig:axial} we show with the dashed line what pion pole dominance predicts if we use the 
fit determined from $G_A(Q^2)$. The error band shows the best fit to $G_p(Q^2)$ if we instead fit the strength
and mass of the monopole in Eq.~\ref{pole dominance}.

\section{N to $\Delta$ transition form factors}
The determination of the N to $\Delta$ electromagnetic and axial transition form factors
requires the
evaluation of the three-point function
$\langle G^{\Delta {\cal O} N}_{\sigma} (t_2, t_1 ;
{\bf p}^{\;\prime}, {\bf p}; \Gamma) \rangle  $ with a new set of inversions,
where for ${\cal O}$ we consider the  electromagnetic and axial currents.
The  $\gamma^* N \Delta$ matrix element is given by
\small
\beq 
\langle\Delta({\bf p}',s')\vert j_\mu\vert N({\bf p},s)\rangle &=&
  i\,\sqrt{\frac{2m_{\Delta}\; m_N}{3E_{\Delta}({\bf p}^\prime)\;E_N({\bf p})}}\bar{u}_\sigma({\bf p}',s')\,\,\,\nonumber \\
& & \hspace*{-4.cm}\Bigg[ G_{M1}(Q^2) K_{\sigma\mu}^{M1} + G_{E2}(Q^2)
  K_{\sigma\mu}^{E2} + G_{C2}(Q^2) K_{\sigma\mu}^{C2} \Bigg]u({\bf p},s) \,.\nonumber
\eeq
\normalsize
The evaluation of the two subdominant electromagnetic form factors $G_{E2}(Q^2)$ and
$G_{C2}(Q^2)$, which are of
primary interest as far as the question of deformation is concerned,
 require high accuracy. A  lattice QCD calculation accurate enough to exclude a zero value
to one standard deviation  would point to  deformation
in the nucleon/$\Delta$ system.
 This is particularly relevant given the fact
 that
extraction of these form factors from experiment involves modeling and
therefore a non-zero value from a first principles calculation
even to one standard deviation is
an important result. Optimized sinks are constructed to isolate the subdominant form 
factors~\cite{Alexandrou:2007dt} along the same lines as discussed for the polarized nucleon matrix element.
In experimental searches for deformation, it is customary to quote the ratios of the electric and Coulomb quadrupole
amplitudes to the magnetic dipole amplitude,  EMR or $ R_{EM} =  -\frac{G_{E2}(Q^2)}{G_{M1}(Q^2)}$$R_{EM}$ 
and CMR or  $R_{SM} = -\frac{\vert\vec{q}\vert}{2m_\Delta}
  \frac{G_{C2}(Q^2)}{G_{M1}(Q^2)}\,,$
in the  rest frame of the $\Delta$. 
\begin{center}
\begin{figure}
\includegraphics[width=7.5cm,height=5cm]{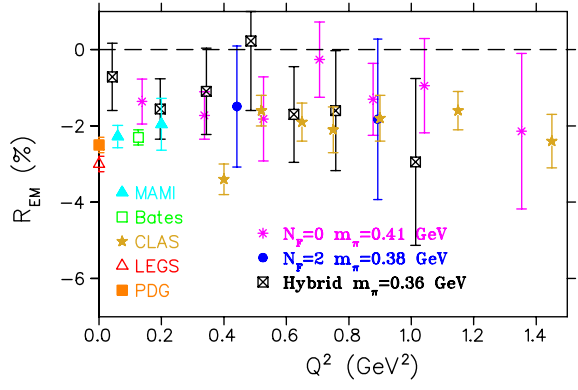}
\caption{\label{fig:EMR} The EMR calculated using quenched (stars) and dynamical (filled circles) Wilson fermions and in the hybrid approach (squares).}
\end{figure}
\end{center}

\begin{center}
\begin{figure}
\includegraphics[width=7.5cm,height=5cm]{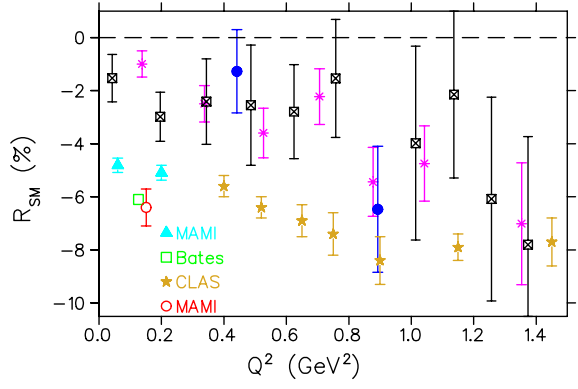}
\caption{\label{fig:CMR} The  CMR calculated using Wilson fermions and in the hybrid approach.}
\end{figure}
\end{center}

Results on these ratios obtained using Wilson fermions and the hybrid action are shown in Figs.~\ref{fig:EMR} 
and \ref{fig:CMR}. Lattice results at low $Q^2$ are non-zero. The lattice values of CMR at small $Q^2$ are less negative
 than experiment. As the pion mas decreases  lattice results tend to become 
more negative approaching experiment.
Therefore one anticipates that for even smaller pion masses the discrepancy between lattice and experiment will be
reduced since pion cloud effects are expected to make CMR more negative as we approach the physical regime~\cite{Pascalutsa:2005}.

 \begin{center}
\begin{figure}
\includegraphics[width=7.5cm,height=5cm]{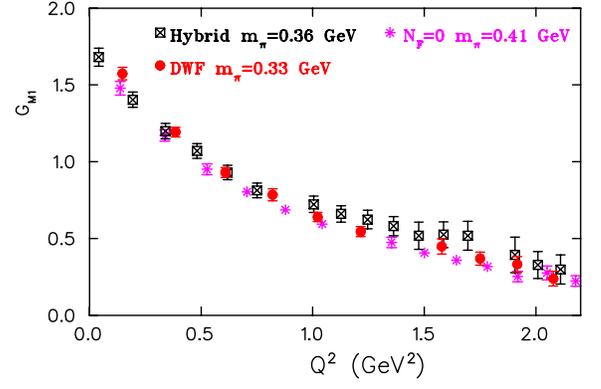}
\caption{\label{fig:GM1} $G_{M1}(Q^2)$
in the hybrid approach and using $N_F=2+1$ domain wall fermions. Quenched Wilson results are also included for comparison.}
\end{figure}
\end{center}
In Fig.~\ref{fig:GM1} we compare results for the dipole form factor  $G_{M1}(Q^2)$ obtained within the hybrid approach and 
using dynamical domain wall fermions at about the same mass. As can be seen there is very good agreement
showing that results obtained  within the non-unitary hybrid action are reliable.

The invariant proton to $\Delta^+$ weak matrix element is expressed
in terms of four transition
form factors as
\small
\beq
<\Delta(p^{\prime},s^\prime)|A^3_{\mu}|N(p,s)> &=& i\sqrt{\frac{2}{3}} 
\left(\frac{m_\Delta m_N}{E_\Delta({\bf p}^\prime) E_N({\bf p})}\right)^{1/2}
\nonumber \\ 
&\>&\hspace*{-4.5cm}\bar{u}_{\Delta^+}^\lambda(p^\prime,s^\prime)\biggl[\left (\frac{C^A_3(q^2)}{m_N}\gamma^\nu + \frac{C^A_4(q^2)}{m^2_N}p{^{\prime \nu}}\right)  
\left(g_{\lambda\mu}g_{\rho\nu}-g_{\lambda\rho}g_{\mu\nu}\right)q^\rho \nonumber \\
&\>&\hspace*{-2cm}+C^A_5(q^2) g_{\lambda\mu} +\frac{C^A_6(q^2)}{m^2_N} q_\lambda q_\mu \biggr]
u_P(p,s)
\label{ND axial}
\eeq
\normalsize
where $C^A_5(Q^2)$ and $C^A_6(Q^2)$, the dominant form factors, can be 
 related assuming pion pole dominance like $G_A(Q^2)$ and $G_p(Q^2)$ are related in the nucleon case.
\begin{center}
\begin{figure}
\includegraphics[width=7.5cm,height=5cm]{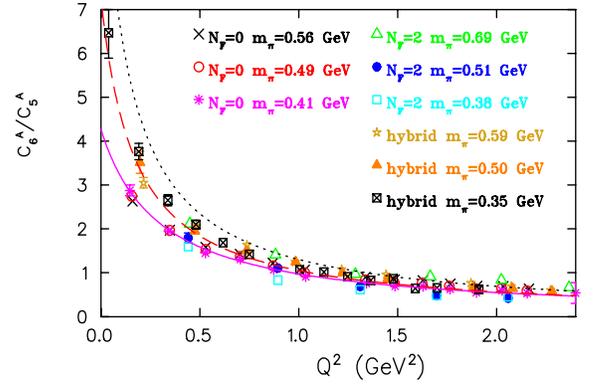}
\caption{\label{fig:CA6overCA5} The ratio of $N$ to $\Delta$ axial transition 
 form factors $C_6^A(Q^2)/C_5^A(Q^2)$. The dotted line  shows
the pion pole dominance prediction for the
hybrid case. The dashed and solid lines are fits to a monopole form for the hybrid and quenched results respectively.}
\end{figure}
\end{center}

In Fig.~\ref{fig:CA6overCA5} we plot the ratio $C^A_6(Q^2)/C^A_5(Q^2)$. The doted line shows
the prediction assuming  pion pole  dominance after a dipole fit to  the hybrid results on $C^A_5(Q^2)$ is performed.
As in the nucleon case the ratio does not fall off as rapidly and a fit to a monopole form to the same hybrid results
gives the dashed line.

\section{$\Delta$ electromagnetic form factors and density distribution}
\begin{center}
\begin{figure}
\includegraphics[width=7.5cm,height=5.5cm]{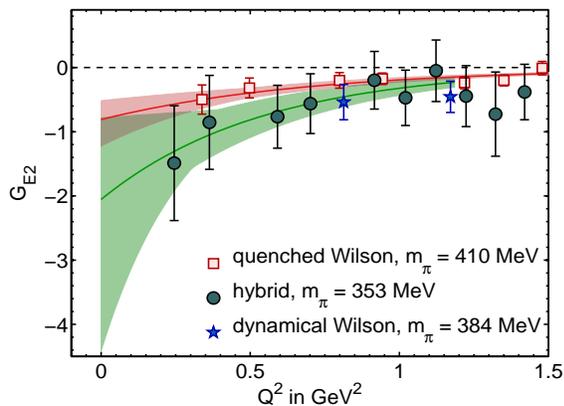}
\caption{\label{fig:GE2} The $Q^2$-dependence of $G_{E2}(Q^2)$. The green (red) 
line and error band show 
a dipole fit to the  mixed action (quenched ) results.
 The value of $G_{E2}$,
in units of $e/(2m_\Delta^2)$, at $Q^2=0$ 
are $-0.810 \pm 291$ for the quenched calculation,  $-0.87 \pm 67$ for $N_F=2$ Wilson
case and $ -2.06^{+1.27}_{-2.35}$ for the hybrid calculation.}
\end{figure}
\end{center}

\begin{center}
\begin{figure}
\includegraphics[width=5.5cm,height=5.5cm]{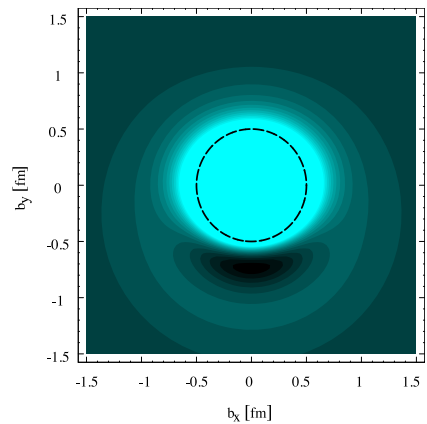}
\caption{\label{fig:deltatrans}Quark transverse charge density in a 
{\it $\Delta^+$} polarized along the $x$-axis, with $s_\perp = +3/2$. 
The light (dark) regions correspond with
the largest (smallest) values of the density.} 
\end{figure}
\end{center}

Since the $\Delta(1232)$ decays strongly,
experiments to measure its form factors are harder than for the N to $\Delta$ transition 
 and yield 
less precise results. The $\Delta$ form factors can be
computed using lattice QCD
more accurately than can be currently obtained from  experiment.
The decomposition for the
on shell $\gamma^* \Delta\Delta$ matrix
element is given by  
\beq
 &\>&\hspace*{-1cm}\langle \Delta(p_f,s_f) | j_{\rm EM}^\mu | \Delta(p_i,s_i)\rangle={\cal A}\>\>\bar{u}_\sigma(p_f,s_f){\cal O}^{\sigma \mu \tau} u_\tau(p_i,s_i) \nonumber \\
&\>&{\cal O}^{\sigma \mu \tau}=-g^{\sigma \tau}\biggl[a_1(q^2) \gamma^\mu +\frac{a_2(q^2)}{2m_\Delta} \left(p_f^\mu + p_i^\mu
\right)\biggr] \nonumber \\
&\>&\hspace*{1.cm}-\frac{q^\sigma q^\tau}{4m_\Delta^2}\biggl[c_1(q^2)\gamma^\mu + \frac{c_2(q^2)}{2m_\Delta}\left(p_f^\mu+p_
i^\mu\right)\biggr] ,
\eeq
 where  $a_1(q^2)$, $a_2(q^2)$, $c_1(q^2)$, and $c_2(q^2)$ are  known linear combinations of   
the electric charge form factor $G_{E0}(q^2)$,
the magnetic dipole form factor $G_{M1}(q^2)$, the electric quadrupole form factor $G_{E2}(q^2)$,
and the magnetic octupole form factor $G_{M3}(q^2)$.
An optimized sink is constructed that isolated the subdominant electric quadrupole form factor~\cite{Alexandrou:2008bn,Alexandrou:2009hs}. The results are shown in Fig.~\ref{fig:GE2}.
The electric quadrupole form factor is particularly interesting because
 it can be related to the shape of a hadron.
Just as the electric form factor for a spin 1/2 nucleon can be 
expressed precisely as the transverse Fourier transform of the 
transverse quark charge density in the infinite momentum frame~\cite{Burkardt:2000za},
 a  proper field-theoretic 
interpretation of the shape of the $\Delta(1232)$ can be 
obtained by considering the quark transverse
charge densities in this frame.
Fig.~\ref{fig:deltatrans} shows the transverse density  
$\rho^\Delta_{T \, s_\perp}$ 
for a $\Delta^+$ with transverse spin $s_\perp = +3/2$ calculated from the fit to the quenched Wilson lattice results for the
 $\Delta$  
form factors (which has the smallest statistical errors of the three calculations). 
It is seen that the $\Delta^+$ quark charge density is elongated 
along the axis of the spin (prolate).   

\section{Conclusions}
Lattice QCD simulations are now being carried out in the chiral regime by a number of collaborations.
We have shown that there is agreement  among recent lattice results using different fermion discretization schemes
 on the low lying baryon spectrum and the nucleon form factors. 
Recent results on the low lying hadron spectrum where
 lattice artifacts have been carefully examined
are in perfect agreement  with experiment providing validation of the lattice 
approach and QCD itself~\cite{Durr:2008zz}. Furthermore we have shown that lattice QCD provides
a framework for the computation of quantities that can not be accurately measured in experiment such as the $\Delta$
form factors providing valuable insight into the structure of such hadrons. We anticipate that other
key hadronic quantities will be computed to sufficient accuracy and with lattice
artifacts taken into account thereby providing direct comparison 
to experiment.

\vspace*{0.5cm}
\acknowledgments{I would like to thank my close collaborators T. Korzec, G. Koutsou, J. W. Negele, C. N. Papanicolas,
E. Stiliaris,  A. Tsapalis and
 M. Vanderhaeghen for their valuable contributions.
I would also like to thank my collaborators in 
 the ETMC for their important contributions in various aspects of the work presented here and for a most enjoyable and
fruitful collaboration. This work is partly supported by 
the Cyprus Research Promotion Foundation under contracts $\Pi$ENEK/ENI$\Sigma$X/0505-39, EPYAN/0506/08 and KY-$\Gamma$A/0907/11.}

\vspace{-2mm}
\centerline{\rule{80mm}{0.1pt}}


\clearpage

\end{document}